# e3D: An Energy-Efficient Routing Algorithm for Wireless Sensor Networks

- Ioan Raicu[♣], Loren Schwiebert[†], Scott Fowler[†], Sandeep K.S. Gupta[‡]

[♣]Department of Computer Science, University of Chicago, iraicu@cs.uchicago.edu
[†]Department of Computer Science, Wayne State University, {loren,fowler}@cs.wayne.edu
[‡]Department of Computer Science and Engineering, Arizona State University, sandeep.gupta@asu.edu

**ABSTRACT**

*One of the limitations of wireless sensor nodes is their inherent limited energy resource. Besides maximizing the lifetime of the sensor node, it is preferable to distribute the energy dissipated throughout the wireless sensor network in order to minimize maintenance and maximize overall system performance. Any communication protocol that involves synchronization of peer nodes incurs some overhead for setting up the communication. We introduce a new algorithm, e3D (energy-efficient Distributed Dynamic Diffusion routing algorithm), and compare it to two other algorithms, namely directed, and random clustering communication. We take into account the setup costs and analyze the energy-efficiency and the useful lifetime of the system. In order to better understand the characteristics of each algorithm and how well e3D really performs, we also compare e3D with its optimum counterpart and an optimum clustering algorithm. The benefit of introducing these ideal algorithms is to show the upper bound on performance at the cost of an astronomical prohibitive synchronization costs. We compare the algorithms in terms of system lifetime, power dissipation distribution, cost of synchronization, and simplicity of the algorithm. Our simulation results show that e3D performs comparable to its optimal counterpart while having significantly less overhead.*

**Keywords:** Simulations, e3D, wireless sensor networks, energy-efficient, routing algorithm, diffusion, clustering

## 1.0 INTRODUCTION

Over the last half a century, computers have exponentially increased in processing power and at the same time decreased in both size and price. These rapid advancements led to a very fast market in which computers would participate in more and more of our society's daily activities. In recent years, one such revolution has been taking place, where computers are becoming so small and so cheap, that single-purpose computers with embedded sensors are almost practical from both economical and theoretical points of view. Wireless sensor networks are beginning to become a reality, and therefore some of the long overlooked limitations have become an important area of research.

In this paper, we attempt to overcome limitations of the wireless sensor networks such as: limited energy resources, varying energy consumption based on location, high cost of transmission, and limited processing capabilities. All of these characteristics of wireless sensor networks are complete opposites of their wired network counterparts, in which energy consumption is not an issue, transmission cost is relatively cheap, and the network nodes have plenty of processing capabilities. Routing approaches that have worked so well in traditional networks for over twenty years will not suffice for this new generation of networks.

Besides maximizing the lifetime of the sensor nodes, it is preferable to distribute the energy dissipated throughout the wireless sensor network in order to minimize maintenance and maximize overall system performance [1, 2]. Any communication protocol that involves synchronization between peer nodes incurs some overhead of setting up the communication. In this paper, we attempt determine whether the benefits of more complex routing algorithms overshadow the extra control messages each node needs to communicate. Each node could make the most informed decision regarding its communication options if they had complete knowledge of the entire network topology and power levels of all the nodes in the network. This indeed proves to yield the best performance if the synchronization messages are not taken into account. However, since all the nodes would always need to have global knowledge, the cost of the synchronization messages would ultimately be very expensive. For both the diffusion and clustering algorithms, we will analyze both realistic and optimum schemes in order to gain more insight in the properties of both approaches.

The usual topology of wireless sensor networks involves having many network nodes dispersed throughout a specific physical area. There is usually no specific architecture or hierarchy in place and therefore, the wireless sensor networks are considered to be ad hoc networks. An ad hoc wireless sensor network may operate in a standalone fashion, or it may be connected to other networks, such as the larger Internet through a base station [3]. Base stations are usually more complex than mere network nodes and usually have an unlimited power supply. Regarding the limited power supply of wireless sensor nodes, spatial reuse of wireless bandwidth, and the nature of radio communication cost which is a function of the distance transmitted squared, it is ideal to send information in several smaller hops rather than one transmission over a long communication distance [4].

In our simulation, we use a data collection problem in which the system is driven by rounds of communication, and each sensor node has a packet to send to the distant base station. The diffusion algorithm is based on location, power levels, and load on the node, and its goal is to distribute the power consumption throughout the network so that the majority of the nodes consume their power supply at relatively the same rate regardless of physical location. This leads to better maintainability of the system, such as replacing the batteries all at once rather than one by one, and maximizing the overall system performance by allowing the network to function at 100% capacity throughout most of its lifetime instead of having a steadily decreasing node population.

## 2.0  SIMULATION RESULTS

Our simulation is based on real world wireless sensors, specifically the Rene RF motes designed at University of California, Berkeley (UCB) [5]. We decided to base our work on these sensors purely because they offer a good architecture to validate the findings of this paper in future work.

In the next few sub-sections, we will discuss the protocols tested in detail. Briefly, the protocols are:

1. Direct communication, in which each node communicates directly with the base station
2. Diffusion-based algorithm utilizing only location data
3. *e*3D: Diffusion based algorithm utilizing location, power levels, and node load
4. An optimum diffusion algorithm using the same metrics as *e*3D, but giving all network nodes global information which they did not have in *e*3D
5. Random clustering, similar to LEACH [6], in which randomly chosen group heads receive messages from all their members and forward them to the base station
6. An optimum clustering algorithm, in which clustering mechanisms are applied at each iteration in order to obtain optimum cluster formation based on physical location and power levels.

Note that the simulation runs presented in sub-sections A to F are all over the same network topology. In order to strengthen our results, we also generated 20 different random network topologies, all containing 100 nodes within a 100 by 100 meter area. The results of the various network topologies were very similar to those presented in this paper and therefore we will not include those results here.

Furthermore, communication medium channel collisions were not simulated, and therefore could affect some of the results. However, considering that for the Rene RF motes the channel capacity is about 25 packets per second, it would seem that collisions would not be a problem if the transmissions would be kept highly localized. Since e3D merely communicates with its close neighbors, collisions are highly unlikely if the interval of transmissions is on the order of seconds. On the other hand, the other routing algorithms presented here would most likely be negatively impacted by communication collisions, and hence perform even worse than they have performed in this paper.

### A. Direct Communication

Each node is assumed to be within communication range of the base station and that they are all aware who the base station is. In the event that the nodes do not know who the base station is, the base station could broadcast a message announcing itself as the base station, after which all nodes in range will send to the specified base station. The simulation assumes that each node transmits at a fixed rate, and always has data to transmit. In every iteration of the simulation, each node sends its data directly to the base station. Eventually, each node will deplete its limited power supply and die. When all nodes are dead, the simulation terminates, and the system is said to be dead. The assumptions stated above will hold for all the algorithm unless otherwise specified.

The main advantages of this algorithm lie in its simplicity. There is no synchronization to be done between peer nodes, and perhaps a simple broadcast message from the base station would suffice in establishing the base station identity. The disadvantages of this algorithm are that radio communication is a function of distance squared, and therefore nodes should opt to transmit a message over several small hops rather than one big one; nodes far away from the base station will die before nodes that are in close proximity of the base station. Another drawback is that communication collision could be a very big problem for even moderate size networks. This can be visualized in Figure 1.

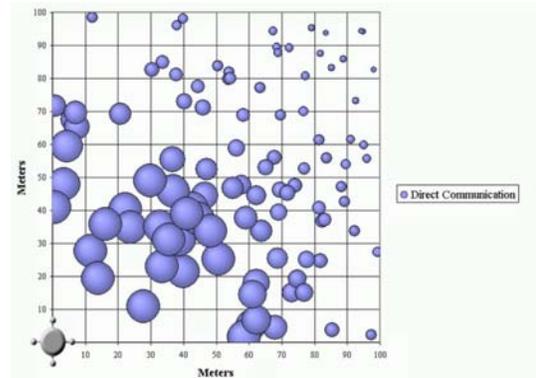
Figure 1: Direct Communication node lifetime

Figures 1-6 all represent the same metric of evaluation, one for each algorithm presented. They follow the same consistency, in which the x axis and y axis represent the physical dimensions of the area while the circles denote a wireless sensor node. The diameter of the circle indicates the relative lifetime of the particular node in relation to other nodes in the network. The bigger the circle, the longer the lifetime of the node was in terms of number of iterations. Obviously, the smaller the circle is, the shorter the lifetime was. The biggest circle had the longest life in the simulation while the smallest circle was the first node that died. The circle with four antennas positioned at coordinates (0,0) is the base station, which is very important to understand the behavior of the various algorithm. The base stations position remained unchanged for the all the simulations and all the algorithms.

### B. Diffusion based algorithm using location information

Each node is assumed to be within communication range of the base station and that they are all aware who the base station is. Once the base station identity is established, the second sequence of messages could be between each node and several of their closest neighbors. Each node is to construct a local table of signal strengths recorded from each of their neighbors, which should be a direct correlation to the distance those nodes are from each other. The other value needed is the distance from each neighbor to the base station, which can be figured out all within the same synchronization messages. This setup phase needs only be completed once at the startup of the system; therefore, it can be considered as constant cost and should not affect the algorithm's performance beyond the setup phase.

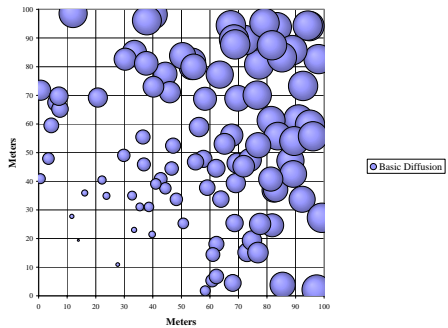

Figure 2: Basic Diffusion node lifetime

The simulation assumes that each node transmits at a fixed rate, and always has data to transmit. In every iteration of the simulation, each node sends its data that is destined for the base station, to the best neighbor. Each node acts as a relay, merely forwarding every message received to its respective neighbor. The best neighbor is calculated using the distance from the sender and the distance from the neighbor to the base station. This ensures that the data is always flowing in the direction of the base station and that no loops are introduced in the system; this can be accomplished by considered not only the distance from the source to the candidate neighbor, but also from the candidate neighbor to the base station. Notice that the complete path is not needed in order to calculate the best optimal neighbor to transmit to. Since each node makes the best decision for itself at a local level, it is inferred that the system should be fairly optimized as a whole.

The main advantage of this system is its fairly light complexity, which allows the synchronization of the neighboring nodes to be done relatively inexpensive, and only once at the system startup. The system also distributes the lifetime of the network a little bit more efficiently.

The disadvantage of this system is that it still does not completely evenly distribute the energy dissipated since nodes close to the base station will die far sooner before nodes far away from the base station. Notice that this phenomenon is inversely proportional to the direct communication algorithm. It should be clear that this happens because the nodes close to the base station end up routing many messages per iteration for the nodes farther away.

### C. *e*3D: Diffusion based algorithm using location, power, and load as metrics

In addition to everything that the basic diffusion algorithm performs, each node makes a list of suitable neighbors and ranks them in order of preference, similar to the previous approach. Every time that a node changes neighbors, the sender will require an acknowledgement for its first message which will ensure that the receiving node is still alive. If a time out occurs, the sending node will choose another neighbor to transmit to and the whole process repeats. Once communication is initiated, there will be no more acknowledgements for any messages. Besides data messages, we introduce exception messages which serve as explicit synchronization messages. Only receivers can issue exception messages, and are primarily used to tell the sending node to stop sending and let the sender choose a different neighbor. An exception message is generated in only three instances: the receiving node's queue is too large, the receiver's power is less than the sender's power, and the receiver has passed a certain threshold which means that it has very little power left.

At any time throughout the system's lifetime, a receiver can tell a sender not to transmit anymore because the receiver's queues are full. This should normally not happen, but in the event it does, an exception message would alleviate the problem. In our current schema, once the sending node receives an exception message and removes his respective neighbor off his neighbor list, the sending node will never consider that same neighbor again. We did this in order to minimize the amount of control messages that would be needed to be exchanged between peer nodes. However, future considerations could be to place a receiving neighbor on probation in the event of an exception message, and only permanently remove it as a valid neighbor after a certain number of exception messages.

The second reason an exception message might be issued, which is the more likely one, is when the receiver's power is less than the sender's power. If we allowed the receiver to send an exception message from the beginning based on this test, most likely the receiver would over-react and tell the sender to stop sending although it is not clear that it was really necessary. We therefore introduced a threshold for the receiver, in which if his own power is less than the specified threshold, it would then analyze the receiving packets for the sender's power levels. If the threshold was made too small, then by the time the receiver managed to react and tell the sender to stop sending, too much of its power supply had been depleted and its life expectancy thereafter would be very limited while the sending node's life expectance would be much longer due to its less energy consumption. Through empirical results, we concluded that the optimum threshold is 50% of the receiver's power levels when it in order to equally distribute the power dissipation throughout the network.

In order to avoid having to acknowledge every message or even have heartbeat messages, we introduce an additional threshold that will tell the receiving node when its battery supply is almost gone. This threshold should be relatively small, in the 5~10% of total power, and is used for telling the senders that their neighbors are almost dead and that new more suitable neighbors should be elected.

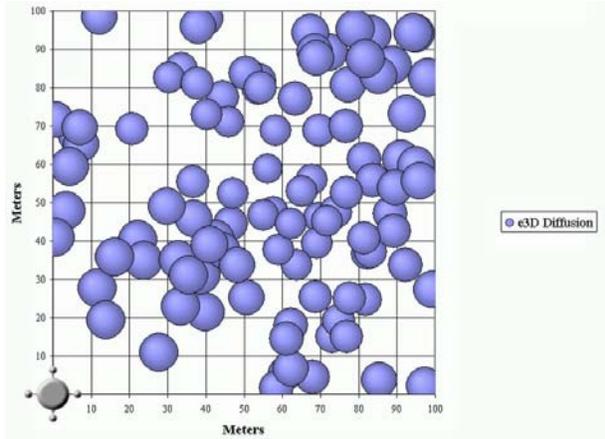

Figure 3: *e*3D Diffusion node lifetime

The synchronization cost of *e*3D is two messages for each pair of neighboring nodes. The rest of the decisions will be based on local look-ups in its memory for the next best suitable neighbor to which it should transmit to. Eventually, when all suitable neighbors are exhausted, the nodes opt to transmit directly to the base station. By looking at the empirical results obtained, it is only towards the end of the system's lifetime that the nodes decide to send directly to the base station.

The main advantage of this algorithm is the near perfect system lifetime where most nodes in the network live relatively the same duration. The system distributes the lifetime and load on the network better than the previous two approaches. The disadvantage when compared to of this algorithm is its higher complexity, which requires some synchronization messages throughout the lifetime of the system. These synchronization message are very few, and therefore worth the price in the event that the application calls for such strict performance.

### D. Ideal Diffusion Based Algorithm

The ideal diffusion based routing algorithm attempts to show the upper bound on performance for diffusion based algorithms. It utilizes all the assumptions and properties of the previous two algorithms, except that all nodes are given global information (power levels and load information) about all other nodes.

Imagine having a directed acyclic tree with the base station as the root. The distance between the nodes times the power levels at the receiver would be the cost for the particular edge. Each node is to find the best neighbor at each iteration, which in principle involves reconstructing the tree at each iteration. Obviously this is almost as hard to achieve in a real world implementation as the clustering techniques we will later discuss, however, the findings here are relevant in order to see the ideal bound on performance for the diffusion based algorithms.

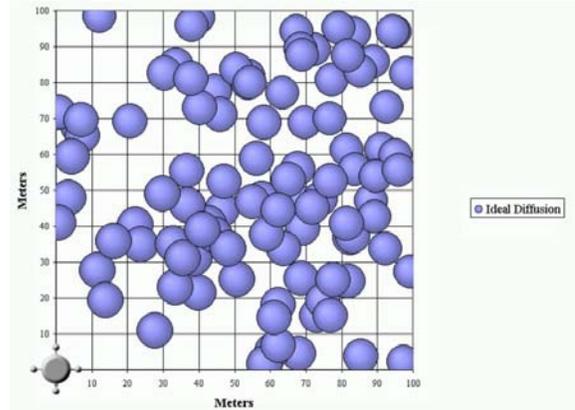

Figure 4: Ideal Diffusion node lifetime

### E. Random Clustering Based Algorithm

This algorithm is similar to LEACH [6], except there is no data aggregation at the cluster heads. Random cluster heads are chosen and clusters of nodes are established which will communicate with the cluster heads.

The main advantage of this algorithm is the distribution of power dissipation achieved by randomly choosing the group heads. This yields a random distribution of node deaths. The disadvantage of this algorithm is its relatively high complexity, which requires many synchronization messages compared to *e*3D at regular intervals throughout the lifetime of the system. Note that cluster heads should not be chosen at every iteration since the cost of synchronization would be very large in comparison to the number of messages that would be actually transmitted. In our simulation, we used rounds of 20 iterations between choosing new cluster heads. The high cost of this schema is not justifiable for the performance gains over much simpler schemes such as direct communication. As a whole, the system does not live very long and has similar characteristics to direct communication, as observed by our simulation in Figure 7. Notice that the only difference in its perceived performance from direct communication is that it randomly kills nodes throughout the network rather than having all the nodes die on one extreme of the network.

Figure 5 shows how nodes with varying distances from the base station died throughout the network. The nodes that are farther away would tend to die earlier because the cluster heads that are farther away have much more work to accomplish than cluster heads that are close to the base station. The random clustering algorithm had a wide range of performance results, which indicated that its performance was directly related to the random cluster election; the worst case scenario had worse performance by a factor of ten in terms of overall system lifetime.

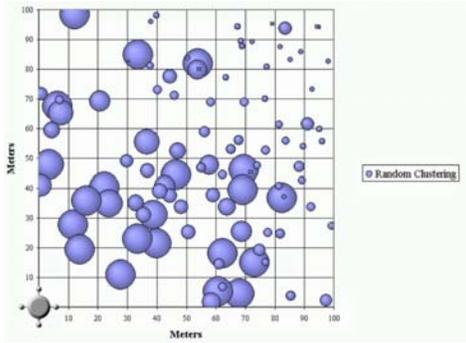

Figure 5: Random Clustering node lifetime

### F. Ideal Clustering Based Algorithm

We implemented this algorithm for comparison purposes to better evaluate the diffusion approach, especially that the random clustering algorithm had a wide range of performance results since everything depended on the random cluster election. The cost of implementing this classical clustering algorithm in a real world distributed system such as wireless sensor networks is energy prohibitively high; however, it does offer us insight into the upper bounds on the performance of clustering based algorithms.

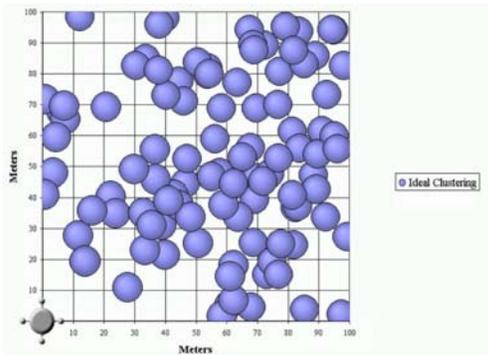

Figure 6: Ideal Clustering node lifetime

We implemented k-Means clustering (k represents the number of clusters) to form the clusters. The cluster heads are chosen to be the clustroid nodes; the clustroid is the node in the cluster that minimizes the sum of the cost metric to the other points of the corresponding cluster. In electing the clustroid, the cost metric is calculated by taking the distance squared between each corresponding node and the candidate clustroid and divided by the candidate clustroid's respective power percentage levels. The metric was calculate at each iteration, and therefore yielded an optimal clustering formation throughout the simulation. We experimented with the number of clusters in order to find the optimum configuration, and discovered that usually between 3 to 10 clusters is optimal for the 20 network topologies we utilized. Notice that the results here are relatively the same as e3D and the ideal diffusion algorithms' results depicted in Figure 3 and Figure 4.

### G. Summary of all the Algorithms

The results for all, but the ideal algorithms include the setup costs and synchronization costs. The cost of synchronization was omitted for the ideal case algorithms because the cost of synchronization would have overshadowed the results; furthermore, the ideal algorithms are not realistic and therefore we only interested on the upper bound they represented.

Figure 7 shows the performance of the system in terms of system lifetime (iterations) and system utility (percentage). Figure 7 and Figure 8 shows that our proposed $e$3D routing algorithm performed almost as good as both ideal diffusion and clustering algorithms. The key idea that needs to be remembered is that the amount of overhead incurred by $e$3D is very minimal and realistic for most applications while both ideal case scenarios are unachievable.

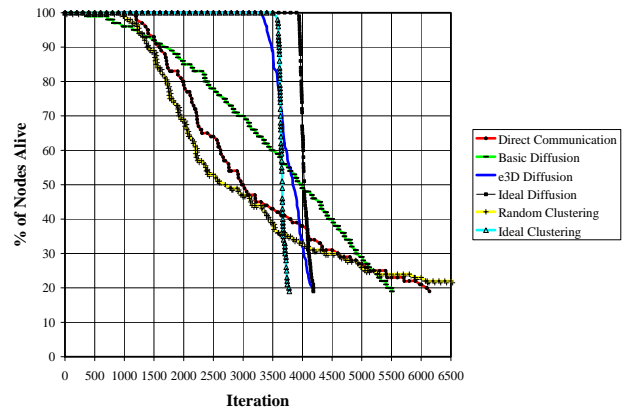

Figure 7: An overview of all the algorithms and their respective lifetime in terms of iterations on average

Figure 7 depicts the network's lifetime in percentage on the y axis and the number of iterations on the x-axis. The six algorithms are all depicted on the same graph in order to easily compare and contrast between the various algorithms. Notice that the random clustering and the direct communication had similar performance. Basic diffusion was a little better, but had an overall similar performance characteristic as direct and random clustering. The remaining three algorithms, $e$3D, the ideal diffusion, and the ideal clustering algorithms, all performed relatively similar. $e$3D was expected to not outperform both ideal cases since it used a realistic scheme for the number of synchronization messages. The ideal diffusion algorithm was also expected to perform better than the ideal clustering since the clustering algorithm cannot avoid sending some message from some nodes backward as they travel from the source to the cluster head and to their final destination at the base station. Since the clustering approach spends more energy in transmitting a message from the source to the destination, the overall system lifetime cannot be expected to be longer than the lifetime represented by the ideal diffusion, in which each source sends the corresponding message along the ideal path towards the base station. Lastly, notice the sharp drop in the percentage of nodes alive, which indicates that the algorithms (*e*3D, ideal diffusion, and ideal clustering) evenly distribute the power dissipated during communication regardless of node location.

Figure 8 attempts to capture an overview comparison between our simulation results (Direct, Diffusion, e3D, Ideal Diffusion, Random Clustering, and Ideal Clustering) and other proposed algorithms (Direct, Leach, Pegasis, MTE, Static Clustering). For the algorithms that are not described in this paper, please refer to [6, 7] for a detailed description.

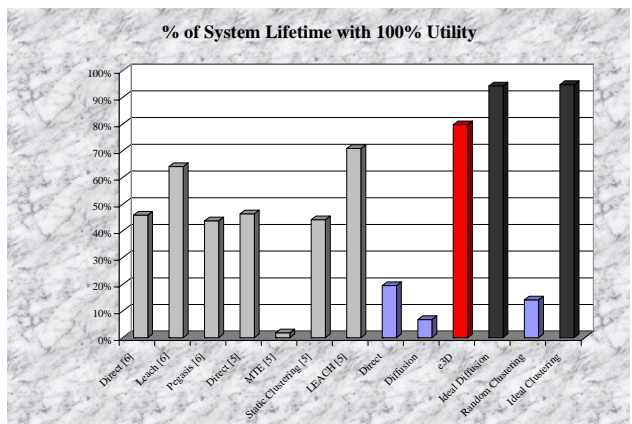

Figure 8: 14 Summary view of algorithms compared in this paper

Figure 8 shows that e3D (red) lived nearly 80% of its lifetime with 100% utility. Some of the related work (Leach) might have had similar system utility (number of iterations while the system had 100% of the nodes alive) because of the use of the unrealistic aggregation scheme which allowed each forwarding node to aggregate unlimited number of incoming packets to one outgoing packets. This in principle placed much less stress on forwarding nodes (cluster heads, neighbors, etc…) and therefore they obtained similar results although under our assumptions, they would have performed much worse. Although no aggregation (data fusion) schemes are used in e3D, it spends almost 80% its system lifetime at 100% system utility, significantly higher than other related work and other algorithms we implemented. Also, note the ideal routing algorithms (black) obtained the expected highest performance spending about 95% of the system lifetime at 100% utility.

## 3. CONCLUSION AND FUTURE WORK

Due to space constraints, we were not able to include all experimental results we have obtained, but we did present the most relevant information to compare *e3D* with other proposed algorithms that had similar goals to ours. The proposed algorithm (*e3D*) performed well in terms of achieving its goal to evenly distribute the power dissipation throughout the network while not creating a very large burden for synchronization purposes.

Our simulation results seem very promising. By distributing the power usage and load on the network, we are essentially improving the quality of the network and making maintenance of it much simpler, since the network lifetime will be predictable as a whole, rather than on a node-by-node basis. In summary, we showed that energy-efficient distributed dynamic diffusion routing is possible at very little overhead cost. The most significant outcome is the near optimal performance of *e*3D when compared to its ideal counterpart in which global knowledge is assumed between the network nodes.

Therefore, we conclude that complex clustering techniques are not necessary in order to achieve good load and power usage balancing. Previous work suggested random clustering as a cheaper alternative to traditional clustering; however, random clustering cannot guarantee good performance according to our simulation results. Perhaps, if aggregation (data fusion) is used, random clustering might be a viable alternative.

Since e3D only addressed static networks, in future work, we will investigate possible modifications so it could support mobility support, and therefore have a wider applicability. We will address the possible aggregation schemes in a future paper in which we discuss in detail both realistic and unrealistic aggregation schemes in order to make the proposed algorithm suitable for most applications. Eventually, it would be nice to implement these algorithms using the Rene RF or MicaZ motes in order to strengthen the simulation results with real world empirical results.